\documentclass[aps,prd,preprint,showpacs,superscriptaddress,tightenlines]{revtex4}
\usepackage{graphics}
\usepackage{tabularx}

\begin{document}
\title{$SU(4)_L\otimes U(1)_X$ three-family model for the electroweak interaction}
\author{Luis A. S\'anchez}
\affiliation{Escuela de F\'\i sica, Universidad Nacional de Colombia,
A.A. 3840, Medell\'\i n, Colombia}
\author{Luis A. Wills-Toro}
\affiliation{Department of Mathematics and Statistics, American University of Sharjah, Po. Box: 26666, Sharjah, United Arab Emirates}
\author{Jorge I. Zuluaga} 
\affiliation{Instituto de F\'\i sica, Universidad de Antioquia,
A.A. 1226, Medell\'\i n, Colombia}


\begin{abstract}
An extension of the gauge group $SU(2)_L\otimes U(1)_Y$ of the standard model to the symmetry group $SU(4)_L\otimes U(1)_X$ (3-4-1 for short) is presented. The model does not contain exotic electric charges and anomaly cancellation is achieved with a family of quarks transforming differently from the other two, thus leading to FCNC. By introducing a discrete $Z_2$ symmetry we obtain a consistent fermion mass spectrum, and avoid unitarity violation of the CKM mixing matrix arising from the mixing of ordinary and exotic quarks. The neutral currents coupled to all neutral vector bosons are studied, and by using CERN LEP and SLAC Linear Collider data at Z-pole and atomic parity violation data, we bound parameters of the model related to tree-level $Z-Z^\prime$ mixing. These parameters are further constrained by using experimental input from neutral meson mixing in the analysis of sources of FCNC present in the model. Constraints coming from the contribution of exotic particles to the one-loop oblique electroweak parameters $S$, $T$ and $U$ are also briefly discussed. Finally, a comparison is done of the predictions of different classes of 3-4-1 models without exotic electric charges.
\end{abstract}

\maketitle

\section{\label{sec:intr}Introduction}
One intriguing puzzle completely unanswered in modern particle physics concerns the number of fermion families in nature. The $SU(3)_c\otimes SU(4)_L\otimes U(1)_X$ extension (3-4-1 for short) of the gauge symmetry $SU(3)_c\otimes SU(2)_L\otimes U(1)_Y$ of the standard model (SM) provides an interesting attempt to answer the question of family replication. In fact, this extension has among its best features that models for the electroweak interaction can be constructed so that anomaly cancellation is achieved by an interplay between generations (three-family models), all of them under the condition that the number of families must be divisible by the number of colors of $SU(3)_c$ \cite{su41,su42}. This happens only if the fermion sector of the model contains an equal number of 4-plets and $4^*$-plets (taking into account the color degree of freedom), which in turn implies, for three families, that one quark multiplet must transform differently from the other two. This fact could give some indication as to why the top quark is unbalancingly heavy, but, at the same time, plagues the model with undesirable flavor changing neutral currents (FCNC) effects. On the other hand, the 3-4-1 gauge structure has been recently considered in order to implement the little Higgs mechanism \cite{little}.

It is well known that the enlargement of the symmetry group of the SM usually leads to fermions in large multiplets having fractionary electric charges different from $\pm 2/3$ and $\pm 1/3$ for exotic quarks, and integer electric charges different from $0$ and $\pm 1$ for exotic leptons, as well as to new gauge bosons with electric charges larger than $1$ and/or fractionary (leptoquarks). A recent analysis of the 3-4-1 gauge theory \cite{sp,sgp} has shown that, by restricting the fermion field representations to particles without exotic electric charges and by paying due attention to anomaly cancellation, only four different three-family models are allowed, while by relaxing the condition of nonexistence of exotic electric charges, an infinite number of models can be constructed.
In this last case, a particular embedding of the SM gauge group into $SU(3)_c\otimes SU(4)_L\otimes U(1)_X$ depends on the physical motivations of the model to be constructed and so, several 3-4-1 models which contain exotic electric charges have been considered in the literature \cite{su41,su42,su4ex}.

The most general expression for the electric charge operator in $SU(4)_L\otimes U(1)_X$ is a linear combination of the four diagonal generators of the gauge group
\begin{equation}\label{ch} 
Q=aT_{3L}+\frac{b}{\sqrt{3}}T_{8L}+
\frac{c}{\sqrt{6}}T_{15L}+ XI_4, \end{equation} 
where $a,b$ and $c$ are free parameters, 
$T_{iL}=\lambda_{iL}/2$, with $\lambda_{iL}$ the Gell-Mann matrices for
$SU(4)_L$ normalized as Tr$(\lambda_i\lambda_j)=2\delta_{ij}$, and
$I_4=Dg(1,1,1,1)$ is the diagonal $4\times 4$ unit matrix. The $X$ values 
are fixed by anomaly cancellation of the fermion content in 
the possible models and an eventual coefficient for $XI_4$ can be absorbed 
in the $X$ hypercharge definition. The free parameters $a,b$ and $c$ fix 
the gauge boson structure of the electroweak sector, and also the electroweak charges of the scalar representations 
which are fully determined by the symmetry breaking pattern implemented. 
In particular $a=1$ gives the usual isospin of the electroweak 
interactions, with $b$ and $c$ remaining as free parameters, producing an 
infinite number of possible models. However, the
restriction to ordinary electric charges, in the fermion, scalar and gauge
boson sectors, allows only for two different cases for the simultaneous
values of the parameters $b$ and $c$, namely: $b = 1, c = -2$ and $b= c = 1$, which become a convenient classification scheme for these type of
models \cite{sp}. Models in the first class differ from those in the second one not only in their fermion content but also in their gauge and scalar boson sectors. As mentioned above, four of the models without exotic electric charges are three-family models. Two of them are models for which $b = 1, c = -2$ and have been partially studied in Refs.~\cite{spp,sedi}. The other two models belong to the class for which $b = c = 1$. One of them has been studied in Ref.~\cite{sgp}, and the other one, the so-called ``Model B" in Ref.~\cite{sp}, has not been yet analyzed in the literature and will be studied in this paper.

This paper is organized as follows. In the next section we review the general conditions to construct anomaly-free 3-4-1 models without exotic electric charges. In Sec.~\ref{sec:sec3} we introduce Model B by describing their anomaly-free fermion content, introduce 
the scalar sector needed to break the symmetry and to produce masses to
the fermion fields, and study the gauge
boson sector paying special attention to the neutral currents and their mixing. In Sec.~\ref{sec:sec4} we analyze the fermion
mass spectrum. In Sec.~\ref{sec:sec5} we use electroweak precision measurements at Z-pole, atomic parity violation (APV) data  and experimental results from FCNC, in order to constrain the mixing angle between two of the neutral currents present in the model and the mass scale of the corresponding new neutral gauge boson. A general discussion about constraints coming from one-loop oblique corrections it is also presented. Sec.~\ref{sec:sec6} is devoted to a comparison between the predictions of the two class of 3-4-1 models without exotic electric charges. In the last section we summarize the model and state our conclusions.
  
\section{\label{sec:sec2}Review of 3-4-1 models}
Let us start with a review of the basics on the construction of 3-4-1 models without exotic electric charges following Ref.~\cite{sp}.

We assume that the electroweak group is 
$SU(4)_L\otimes U(1)_X\supset SU(3)_L \otimes U(1)_Z 
\supset SU(2)_L \otimes U(1)_Y$, where the gauge structure 
$SU(3)_L\otimes U(1)_Z$ refers to the one presented in Ref.~\cite{331}. 
We also assume that the left handed quarks and left-handed 
leptons transform either under the 4 or 
the $4^*$ fundamental representations of $SU(4)_L$, and that all the right-handed charged fermions transform as $SU(4)_L$ singlets. As in the SM, $SU(3)_c$ is vectorlike.

As stated in the introduction, in $SU(4)_L\otimes U(1)_X$ the most general expression for the electric charge generator is given by Eq.~(\ref{ch}). If we assume that the usual isospin $SU(2)_L$ of the SM is such that
$SU(2)_L\subset SU(4)_L$, and we demand for accomodating each family of SM fermions into different fundamental representations $4$ or $4^*$ of $SU(4)_L$, then $a=1$ and we have just a two-parameter set
of models, all of them characterized by the values of $b$ and $c$. So, 
Eq.~(\ref{ch}) allows for an infinite number of models in the context of the 3-4-1 theory, each one associated to particular values of the parameters $b$ and $c$, with characteristic signatures that make them different from each other.

There are a total of 24 gauge bosons in the gauge group under
consideration: the 8 gluon fields associated with $SU(3)_c$, one gauge field $B^\mu$ associated with $U(1)_X$, and another 15 gauge bosons associated with $SU(4)_L$ which we may write as
\begin{widetext}
\begin{equation}
\frac{1}{2}\lambda_\alpha A^\alpha_\mu =\frac{1}{\sqrt{2}} 
\left(\begin{array}{cccc}D^0_{1\mu} & W^+_\mu & K^{(b+1)/2}_\mu & X^{(3+b
+2c)/6}_\mu\\ 
W^-_\mu & D^0_{2\mu} &  K^{(b-1)/2}_{1\mu} &  V^{(-3+b+2c)/6}_\mu \\
K^{-(b+1)/2}_\mu & K^{-(b-1)/2}_{1\mu} & D^0_{3\mu} & Y^{-(b-c)/3}_\mu\\
X^{-(3+b+2c)/6}_\mu & V^{(3-b-2c)/6}_\mu & Y^{(b-c)/3}_\mu & 
D^0_{4\mu} \end{array}\right),  
\label{gauge}
\end{equation}
\end{widetext}
where $D^{0\mu}_1=A_3^\mu/\sqrt{2}+A_8^\mu/\sqrt{6}+A_{15}^\mu/\sqrt{12}$;
$D^{0\mu}_2=-A_3^\mu/\sqrt{2}+A_8^\mu/\sqrt{6}+A_{15}^\mu/\sqrt{12}$;
$D^{0\mu}_3=-2A_8^\mu/\sqrt{6}+A_{15}^\mu/\sqrt{12}$, and 
$D^{0\mu}_4=-3 A_{15}^\mu/\sqrt{12}$. The upper indices in the gauge bosons in the former expression stand for the electric charge of the corresponding 
particle, some of them functions of the $b$ and $c$ parameters as they 
should be. 

Different from the SM where only the abelian $U(1)_Y$ factor is
anomalous, in the 3-4-1 theory both, $SU(4)_L$ and $U(1)_X$ are anomalous. So, special combinations of multiplets must be
used in each particular model in order to cancel the possible
anomalies, and obtain renormalizable models. The triangle anomalies
we must take care of are: $[SU(4)_L]^3$, $[SU(3)_c]^2U(1)_X$, 
$[SU(4)_L]^2U(1)_X$, $[grav]^2U(1)_X$ and $[U(1)_X]^3$.

\subsection{\label{sec:sub2a}Models without exotic electric charges}

Let us now see how the charge operator in Eq.~(\ref{ch}) acts on the
representations 4 and $4^*$ of $SU(4)_L$:
\begin{widetext}
\begin{eqnarray}\nonumber
Q[4]
&=&Dg.\left(\frac{1}{2}+\frac{b}{6}+\frac{c}{12}+X, -\frac{1}{2}+\frac{b}{6}+\frac{c}{12}+X, -\frac{2b}{6}+\frac{c}{12}+X, -\frac{3c}{12}+X\right),\\ \label{eq3}
Q[4^*]
&=&Dg.\left(-\frac{1}{2}-\frac{b}{6}-\frac{c}{12}+X, \frac{1}{2}-\frac{b}{6}-\frac{c}{12}+X, \frac{2b}{6}-\frac{c}{12}+X, \frac{3c}{12}+X\right).
\end{eqnarray}
\end{widetext}

Notice, from Eq.~(\ref{eq3}), that if we demand for gauge bosons with electric charges $0, \pm 1$ only, there are not more than four different possibilities for the simultaneous values of $b$ and $c$; they are: 
$b= c = 1$; $b = c = -1$; $b = 1$, $c = -2$, and $b = -1$, $c = 2$.
 
It is clear that, if we accommodate the known
left-handed quark and lepton isodoublets in the two upper components of 4
and $4^*$ (or $4^*$ and 4), do not allow for electrically charged
antiparticles in the two lower components of the multiplets (antiquarks
violate $SU(3)_c$, and $e^+, \mu^+$ and $\tau^+$ violate lepton number at
tree level) and forbid the presence of exotic electric charges in the
possible models, then the electric charge of the third and fourth
components in $4$ and $4^*$ must be equal either to the charge of the
first and/or second component, which in turn implies that $b$ and $c$ can
take only the four sets of values stated above. So, these four sets of
values for $b$ and $c$ are necessary and sufficient conditions in order to
exclude exotic electric charges in the fermion sector too.

A further analysis also shows that models with $b=c=-1$ are equivalent,
via charge conjugation, to models with $b=c=1$. Similarly, models 
with $b=-1,\; c=2$ are equivalent to models with $b=1,\; c=-2$. So, 
with the constraints impossed, we have only two different sets of models: 
those for $b=c=1$ and those for $b=1, \; c=-2$. 

\subsubsection{\label{sec:sub2a1}Models for $b=c=1$}
Let us start defining the following complete sets of spin 1/2 Weyl spinors 
(complete in the sense that each set contains its own charged 
antiparticles):
\begin{eqnarray}\nonumber
S_1^q&=& \{(u,d,D,D^\prime)_L\sim [3,4,-\frac{1}{12}], \; 
u_L^c \sim [3^*, 1, -\frac{2}{3}], \nonumber \\ 
& & d_L^c \sim [3^*,1,\frac{1}{3}], \; 
D_L^c \sim [3^*, 1, \frac{1}{3}], \; 
D_L^{\prime c} \sim [3^*, 1, \frac{1}{3}]\}. \nonumber \\ 
S_2^q&=& \{(d,u,U,U^\prime)_L\sim [3,4^*,\frac{5}{12}], \; 
u_L^c \sim [3^*, 1, -\frac{2}{3}], \nonumber \\ 
& & d_L^c \sim [3^*,1,\frac{1}{3}], \; 
U_L^c \sim [3^*, 1, -\frac{2}{3}], \; 
U_L^{\prime c} \sim [3^*,1,-\frac{2}{3}]\}. \nonumber \\ 
S_3^l&=& \{(\nu^0_e, e^-, E^-, E^{\prime -})_L \sim 
[1,4,-\frac{3}{4}],\;
e^+_L \sim [1,1,1], \nonumber \\
& & E^+_L \sim [1,1,1], \; E^{\prime +}_L \sim 
[1,1,1]\}. \nonumber \\ 
S_4^l&=& \{(E^+, N^0_1, N^0_2, N^0_3)_L\sim 
[1, 4, \frac{1}{4}], \; E^-_L\sim [1,1,-1]\}. \nonumber \\ 
S_5^l &=& \{(e^-, \nu^0_e, N^0, N^{\prime 0})_L\sim 
[1, 4^*, -\frac{1}{4}], \; e^+_L\sim [1,1,1]\}. \nonumber \\
S_6^l&=& \{(N^0, E_1^+, E_2^+, E_3^+)_L \sim [1, 4^*, 
\frac{3}{4}],\; E_{1L}^- \sim [1,1,-1], \nonumber \\ 
& & E_{2L}^- \sim [1,1,-1], \; E_{3L}^- \sim 
[1,1,-1]\}. \nonumber
\end{eqnarray}
The numbers in parentheses refer to the $[SU(3)_C, SU(4)_L, U(1)_X]$ quantum numbers respectively.

Taking into account that each set includes charged particles together with
their corresponding antiparticles, and since $SU(3)_c$ is vectorlike, the
anomalies $[grav]^2U(1)_X, \; [SU(3)_c]^3$ and $[SU(3)_c]^2U(1)_X$
automatically vanish. So, we only have to take care of the remaining three
anomalies whose values are shown in Table~\ref{tab1}, from which only two different anomaly free three-famlily models can be constructed:

Model A = $2S_1^q \oplus S_2^q \oplus 3S_5^l$. (This model 
has been analyzed in Ref.~\cite{sgp}). 

Model B = $S_1^q \oplus 2S_2^q \oplus 3S_3^l$. 

These two fermion structures show that anomaly cancellation between generations forces one family of quarks to transform different from the other two.

\begin{table*}
\caption{\label{tab1}Anomalies for models with $b$=$c$=1}
\begin{ruledtabular}
\begin{tabular}{lcccccc} 
Anomaly & $S_1^q$ & $S_2^q$ & $S_3^l$ & $S_4^l$ & $S_5^l$ & $S_6^l$ \\ 
\hline
$[U(1)_X]^3$ & $-9/16$ & $-27/16$ & 21/16 & $-15/16$ & 15/16 & $-21/16$ \\
$[SU(4)_L]^2U(1)_X$ & $-1/4$ & 5/4 & $-3/4$ & 1/4 & $-1/4$ & 3/4 \\
$[SU(4)_L]^3 $ & 3 & $-3$ & 1 & 1 & $-1$ & $-1$ \\  
\end{tabular}
\end{ruledtabular}
\end{table*}

\begin{table*}
\caption{\label{tab2}Anomalies for models with $b=1, \; c=-2$}
\begin{ruledtabular}
\begin{tabular}{lcccccc} 
Anomaly & $S_1^{\prime q}$ & $S_2^{\prime q}$ & $S_3^{\prime l}$ & $S_4^{\prime l}$ & $S_5^{\prime l}$ & $S_6^{\prime l}$ \\ 
\hline
$[U(1)_X]^3$ & $-3/2$ & $-3/2$ & 3/2 & $ 3/2 $ & -3/2 & $-3/2 $ \\
$[SU(4)_L]^2U(1)_X$ & 1/2 & 1/2 & $-1/2$ & $-1/2$ & 1/2 & 1/2 \\
$[SU(4)_L]^3 $ & 3 & $-3$ & 1 & $-1$ & 1 & $-1$  \\ 
\end{tabular}
\end{ruledtabular}
\end{table*}

\subsubsection{\label{sec:sub2a2}Models for $b=1, c=-2$}
As in the previous case, let us define the following complete sets of spin 
1/2 Weyl spinors:
\begin{eqnarray}
S_1^{\prime q}&=& \{(u,d,D,U)_L\sim [3,4,\frac{1}{6}], \; 
u_L^c \sim [3^*, 1, -\frac{2}{3}], \nonumber \\ 
& & d_L^c \sim [3^*,1,\frac{1}{3}], \; 
D_L^c \sim [3^*, 1, \frac{1}{3}], \; 
U_L^{c} \sim [3^*, 1, -\frac{2}{3}]\}. \nonumber \\ 
S_2^{\prime q}&=& \{(d,u,U,D)_L\sim [3,4^*,\frac{1}{6}], \; 
u_L^c \sim [3^*, 1, -\frac{2}{3}], \nonumber \\  
& & d_L^c \sim [3^*,1,\frac{1}{3}], \; 
U_L^c \sim [3^*, 1, -\frac{2}{3}], \; 
D_L^{c} \sim [3^*,1,\frac{1}{3}]\}. \nonumber \\  
S_3^{\prime l}&=& \{(\nu^0_e, e^-, E^-, N^0)_L \sim 
[1,4,-\frac{1}{2}],\;
e^+_L \sim [1,1,1], \nonumber \\ 
& & E^+_L \sim [1,1,1]\}. \nonumber \\ 
S_4^{\prime l} &=& \{(e^-, \nu^0_e, N^0, E^-)_L\sim 
[1, 4^*, -\frac{1}{2}], \; e^+_L\sim [1,1,1], \nonumber \\ 
& & E^+_L\sim [1,1,1]\}. \nonumber \\  
S_5^{\prime l} &=& \{(E^+, N^0_1, N^0_2, e^+)_L\sim 
[1, 4, \frac{1}{2}], \; E^-_L\sim [1,1,-1], \nonumber \\  
& & e^-_L\sim [1,1,-1]\}. \nonumber \\  
S_6^{\prime l}&=& \{(N_3^0, E^+, e^+, N_4^0)_L \sim 
[1, 4^*, \frac{1}{2}],\;
E_L^- \sim [1,1,-1], \nonumber \\ 
& & e_{L}^- \sim [1,1,-1]\}. \nonumber
\end{eqnarray}

For these sets the anomalies $[grav]^2U(1)_X, \; [SU(3)_c]^3$ and
$[SU(3)_c]^2U(1)_X$ vanish.  The other anomalies are shown in 
Table~\ref{tab2} from which the following two different anomaly free three-family structures can be extracted:

Model E = $2S_1^{\prime q} \oplus S_2^{\prime q} \oplus 
3S_4^{\prime l}$. (This model has been studied in Ref.~\cite{spp}).
 
Model F = $S_1^{\prime q} \oplus 2S_2^{\prime q} \oplus 
3S_3^{\prime l}$. (This model has been studied in Ref.~\cite{sedi})

Again, anomaly cancellation between the families implies one family of quarks transforming different from the other two under the gauge group.
 
\section{\label{sec:sec3}Model B}
Model B belongs to the class for which $b=c=1$. In this case the electric charge generator in Eq.~(\ref{ch}) reads: $Q=T_{3L}+T_{8L}/\sqrt{3}+T_{15L}/\sqrt{6}+ XI_4$, and the gauge bosons in $SU(4)_L$ are obtained from Eq.~(\ref{gauge}) as
\[\frac{1}{2}\lambda_{L\alpha} A^\alpha_\mu=\frac{1}{\sqrt{2}}\left(
\begin{array}{cccc}D^0_{1\mu} & W^{+}_\mu & K^{+}_\mu & X^{+}_\mu\\ 
W^{-}_\mu & D^{0}_{2\mu} &  K^{0}_{\mu} &  X^{0}_\mu\\
K^{-}_\mu & K^{\prime 0}_{\mu} & D^{0}_{3\mu} & Y^{0}_\mu\\
X^{-}_\mu & X^{\prime 0}_\mu & Y^{\prime 0}_\mu & D^{0}_{4\mu} \end{array}\right),\]
where we have renamed $V_\mu$ as $X^0_\mu$.

The gauge couplings $g_4$ and $g_X$, associated with the groups $SU(4)_L$ and $U(1)_X$, respectively, are defined through the covariant derivative for 4-plets as: $iD^\mu=i\partial^\mu -g_4 \lambda_{L\alpha} A^\mu_\alpha/2-g_XXB^\mu$. 

\subsection{\label{sec:sub3a}The spin $1/2$ particle representations}
For Model B the anomaly-free fermion content is given by $S_1^q \oplus 2S_2^q \oplus 3S_3^l$, which is displayed as
\begin{eqnarray*}
Q_{iL}&=&\left(\begin{array}{c}d_i\\u_i\\U_i\\U^\prime_i 
\end{array}\right)_L \sim [3,4^*,{5\over 12}], \\
d^c_{iL}&\sim & [3^*,1,{1\over 3}],\quad u^c_{iL} \sim [3^*,1,-{2\over 3}],\\
\quad U^c_{iL}&\sim &[3^*,1,-{2\over 3}], \quad U^{'c}_{iL}\sim [3^*,1,-{2\over 3}],
\end{eqnarray*}

\begin{eqnarray*}
Q_{3L}&=&\left(\begin{array}{c}u_3\\d_3\\D_3\\D'_3 \end{array}\right)_{L}  \sim [3,4,-{1\over 12}], \\
u^c_{3L}&\sim &[3^*,1,-{2\over 3}],\quad d^c_{3L} \sim [3^*,1,{1\over 3}],\\
\quad D^c_{3L}&\sim &[3^*,1,{1\over 3}], \quad D^{'c}_{3L}\sim [3^*,1,{1\over 3}],
\end{eqnarray*}

\begin{eqnarray*}
L_{\alpha L}&=&\left(\begin{array}{c} \nu^0_{e \alpha}\\ e^-_{\alpha}\\ 
E^-_\alpha \\ E^{'-}_\alpha \end{array}\right)_L\sim [1,4,-{3\over 4}], \\ 
e^+_{\alpha L}&\sim &[1,1,1],\quad E^+_{\alpha L}\sim [1,1,1],\quad E^{'+}_{\alpha L}\sim [1,1,1],
\end{eqnarray*}
\noindent
where $i=1,2$ and $\alpha = 1,2,3$ are family indexes. Notice that anomaly cancellation among the families implies a non-universal hypercharge $X$ for the left-handed quark multiplets.

The former expressions for the quark multiplets are written in the weak basis, in which the distinction between quark families is meaningless. However, when we go to mass eigenstates, the almost diagonal Cabibbo-Kobayashi-Maskawa (CKM) mixing matrix suggest us to classify them in families. In this sense we have, in principle, three different assignments of weak eigenstates into mass eigenstates. We will assign the heaviest family of quarks to the $4$-plet. As we will discuss ahead, this choice makes difference when the phenomenological implications of the model are examined. Notice also that universality for the known leptons in the three families is present at tree level in the weak basis. As a result, FCNC do not occur in the lepton sector, up to possible mixing with the exotic fields. 
\subsection{\label{sec:sub3b}Scalars}
In order to have at the electroweak scale ($v_{EW} = 174$~GeV), an effective theory resembling the two Higgs doublet extension of the SM (THDM) \cite{thdm1}, to avoid unnecesary mixing in the electroweak gauge boson sector, and to 
give masses to the fermion fields in the model, we introduce the following four Higgs scalars, and vacuum expectation values (VEV) aligned as:
\begin{eqnarray}\nonumber 
\langle\phi^T_1\rangle&=&\langle(\phi^0_1,\phi^+_1,\phi^{\prime +}_1,\phi^{\prime\prime +}_1)\rangle=(v,0,0,0)\sim[1,4^*,3/4], \\ \nonumber
\langle\phi^T_2\rangle&=&\langle(\phi^-_2,\phi^0_2,\phi^{\prime 0}_2,\phi^{\prime\prime 0}_2)\rangle=(0,v',0,0)\sim[1,4^*,-1/4], \\ \nonumber
\langle\phi^T_3\rangle&=&\langle(\phi^-_3,\phi^0_3,\phi^{\prime 0}_3,\phi^{\prime\prime 0}_3)\rangle=(0,0,V,0)\sim[1,4^*,-1/4], \\ \nonumber
\langle\phi^T_4\rangle&=&\langle(\phi^-_4,\phi^0_4,\phi^{\prime 0}_4,\phi^{\prime\prime 0}_4)\rangle=(0,0,0,V')\sim[1,4^*,-1/4], \\ \label{scalars}
\end{eqnarray}
where we assume the hierarchy $V\sim V^\prime >> v\sim v^{\prime} \simeq 174$~GeV.

This set of scalars break the symmetry in three steps
\begin{eqnarray}\nonumber
SU(4)_L\otimes & U(1)_X & \\ \nonumber
& \stackrel{V^\prime}{\longrightarrow}
& SU(3)_L\otimes U(1)_Z \\ \nonumber 
& \stackrel{V}{\longrightarrow} & SU(2)_L\otimes U(1)_Y \\ \label{break}
& \stackrel{v+v^\prime}{\longrightarrow} & U(1)_Q.
\end{eqnarray}

From the charge operator we have that the hypercharge of the SM is given by $Y/2=T_{8L}/\sqrt{3}+T_{15L}/\sqrt{6}+X$, thus, when the 3-4-1 symmetry is broken to the SM, we obtain the gauge matching conditions

\begin{eqnarray}
g_4 = g, \qquad&\mbox{and}&\qquad\frac{1}{g^{\prime 2}}=\frac{1}{g^2_X}+\frac{1}{2g^2}, \label{match}
\end{eqnarray}
where $g$ and $g^\prime$ are the gauge coupling constants of the $SU(2)_L$ and $U(1)_Y$ gauge groups of the SM, respectively.

\subsection{\label{sec:sub3c}Gauge boson masses and currents}

After the 3-4-1 symmetry is broken to $SU(3)_c\otimes U(1)_Q$, we obtain that the charged gauge bosons (charged under the generators of the $SU(4)_L$ group) do not mix with each other and get the following squared masses
\begin{eqnarray}\nonumber
M^2_{W^\pm}&=&\frac{g_4^2}{2}(v^2+v^{\prime 2}), \quad M^2_{K^\pm}=\frac{g_4^2}{2}(v^2+V^2), \\ \nonumber
M^2_{X^\pm}&=&\frac{g_4^2}{2}(v^2+V^{\prime 2}), \quad
M^2_{K^0(K^{\prime 0})}=\frac{g_4^2}{2}(v^{\prime 2}+V^2), \\ \nonumber
M^2_{X^0(X^{\prime 0})}&=&\frac{g_4^2}{2}(v^{\prime 2}+V^{\prime 2}), \quad M^2_{Y^0(Y^{\prime 0})}=\frac{g_4^2}{2}(V^2+V^{\prime 2}). \\ \label{chmass}
\end{eqnarray}
Notice that $W^\pm$ does not mix with the other charged bosons. Then, with $g_4=g$ and the experimental value $M_W=80.450\pm 0.058$ GeV \cite{pdg}, we have that $\sqrt{v^2+v^{\prime 2}}\approx 174$ GeV.

For the four neutral gauge bosons we get mass terms of the form 

\begin{eqnarray*}
& &{g_4^2\over 2}\Big\{V^2 \left(\frac{g_XB^\mu}{2g_4}
-\frac{2A_8^\mu}{\sqrt{3}}+\frac{A^\mu_{15}}{\sqrt{6}}\right)^2 \\
& & + V^{\prime 2}\left(\frac{g_XB^\mu}{2g_4}-\frac{3 
A^\mu_{15}}{\sqrt{6}}\right)^2 \\
& &+v^{\prime 2}\left(\frac{g_XB^\mu}{2g_4}-A^\mu_3
+\frac{A_8^\mu}{\sqrt{3}}+\frac{A^\mu_{15}}{\sqrt{6}}
\right)^2 \\
& & + v^2 \left(\frac{-3g_XB^\mu}{2g_4}+A^\mu_3
+\frac{A_8^\mu}{\sqrt{3}}+\frac{A^\mu_{15}}{\sqrt{6}}\right)^2 \Big\},
\end{eqnarray*}
\noindent 
associated to a $4 \times 4$ mass matrix with a zero eigenvalue
corresponding to the photon. Once the photon field has been identified, we
remain with a $3 \times 3$ mass matrix for three neutral gauge bosons
$Z^\mu$, $Z^{'\mu}$ and $Z^{''\mu}$. Since we are interested in 
the low energy phenomenology of the model, we can choose $V=V^\prime$ in order to simplify matters. Also, the mixing between the three neutral gauge bosons can be further simplified by choosing $v^\prime = v$. In this case the field $Z''^\mu= A_8^\mu / \sqrt{3}-\sqrt{2/3}A_{15}^\mu$ decouples from the other two and acquires a squared mass $(g_4^2/2)V^2$. The remaining $2 \times 2$ mass matrix, in the basis $(Z_\mu,Z^{\prime}_\mu)$, is

\[\frac{g_4^2}{C^2_W}\left(\begin{array}{cc} v^2 & \sqrt{2}\delta v^2 S_W \\
\sqrt{2}\delta v^2 S_W & \frac{2 \delta^2}{S^2_W}[v^2(S^4_W+C^4_W)+V^2 C^4_W]\end{array}\right),\]
\noindent
where $\delta=g_X/(2g_4)$, and $S_W=2 \delta /\sqrt{6 \delta^2 + 1}$ and $C_W$ are the sine and cosine of the electroweak mixing angle, respectively. 

By diagonalizing this mass matrix we get the two physical neutral gauge bosons
\begin{eqnarray}\nonumber
Z_1^\mu&=&Z^\mu \cos\theta+Z^{\prime\mu} \sin\theta \; ,\\ \label{mixing}
Z_2^\mu&=&-Z^\mu \sin\theta+Z^{\prime\mu} \cos\theta, 
\end{eqnarray} 
where the mixing angle is given by
\begin{equation} \label{tan} \tan(2\theta) = \frac{2 \sqrt{2} \delta v^2 S^3_W}
{2 \delta^2[v^2(S^4_W+C^4_W)+V^2 C^4_W]-v^2 S^2_W}. 
\end{equation}
\noindent  
The photon field $A^\mu$ and the fields $Z_\mu$ and $Z'_\mu$ are given by

\begin{eqnarray} \nonumber
A^\mu&=&S_W A_3^\mu + C_W Y^\mu\; , \\ \nonumber  
Z^\mu&=& C_W A_3^\mu - S_W Y^\mu \; , \\ \nonumber 
Z'^\mu&=&\sqrt{\frac{2}{3}}(1-T_W^2/2)^{1/2}\left(A_8^\mu+
\frac{A_{15}^\mu}{\sqrt{2}}\right)-\frac{T_W}{\sqrt{2}}B^\mu, \\ \label{fzzp}
\end{eqnarray}
\noindent
where $T_W=S_W/C_W$, and
\begin{equation}\label{y}
Y^\mu=\frac{T_W}{\sqrt{3}}\left(A_8^\mu+
\frac{A_{15}^\mu}{\sqrt{2}}\right)+(1-T_W^2/2)^{1/2}B^\mu,
\end{equation}
is the field to be identified as the $Y$ hypercharge associated with the SM abelian gauge boson.
\subsubsection{\label{sec:sub3c1}Charged currents}
The Lagrangian for the charged currents is $-{\cal L}^{CC}=g_4(W^+_\mu J^\mu_{W^+}+K^+_\mu J^\mu_{K^+} +X^+_\mu J^\mu_{X^+}+K^0_\mu J^\mu_{K^0}+X^0_\mu J^\mu_{X^0}+Y^0_\mu J^\mu_{Y^0})/\sqrt{2}+\mbox{h.c.}$, where
\begin{widetext}
\begin{eqnarray}\nonumber
J^\mu_{W^+} &=&\bar{u}_{3L}\gamma^\mu d_{3L} - \sum_{i=1}^2\bar{u}_{iL}\gamma^\mu d_{iL} + \sum_{\alpha=1}^3
\bar{\nu}_{e\alpha L}\gamma^\mu e^-_{\alpha L}, \\ \nonumber 
J^\mu_{K^+} &=&\bar{u}_{3L}\gamma^\mu 
D_{3L} - \sum_{i=1}^2\bar{U}_{iL}\gamma^\mu 
d_{iL} + \sum_{\alpha=1}^3
\bar{\nu}_{e\alpha L}\gamma^\mu E^-_{\alpha L}, \\ \nonumber 
J^\mu_{X^+} &=& \bar{u}_{3L}\gamma^\mu 
D'_{3L} -\sum_{i=1}^2\bar{U'}_{iL}\gamma^\mu 
d_{iL} + \sum_{\alpha=1}^3
\bar{\nu}_{e\alpha L}\gamma^\mu E^{\prime -}_{\alpha L}, \\ \nonumber 
J^\mu_{K^0} &=& \bar{d}_{3L}\gamma^\mu 
D_{3L} - \sum_{i=1}^2\bar{U}_{iL}\gamma^\mu 
u_{iL} + \sum_{\alpha=1}^3
\bar{e}^-_{\alpha L}\gamma^\mu E^-_{\alpha L}, \\ \nonumber 
J^\mu_{X^0} &=& \bar{d}_{3L}\gamma^\mu 
D'_{3L} - \sum_{i=1}^2\bar{U'}_{iL}\gamma^\mu 
u_{iL} + \sum_{\alpha=1}^3
\bar{e}^-_{\alpha L}\gamma^\mu E^{\prime -}_{\alpha L}, \\  
J^\mu_{Y^0} &=& \bar{D}_{3L}\gamma^\mu 
D'_{3L} - \sum_{i=1}^2\bar{U'}_{iL}\gamma^\mu 
U_{iL} + \sum_{\alpha=1}^3
\bar{E}^-_{\alpha L}\gamma^\mu E^{\prime -}_{\alpha L}.
\end{eqnarray}
\end{widetext}
\subsubsection{\label{sec:sub3c2}Neutral currents}
The Lagrangian for neutral currents can be written as 
$-{\cal L}^{NC} = eA^\mu J_\mu(EM)+(g_4/{C_W})Z^\mu J_\mu(Z)
+ (g_X/\sqrt{2})Z'^\mu J_\mu(Z') + (g_4/2)Z''^\mu J_\mu(Z'')$,
with
\begin{eqnarray}\nonumber
J_\mu(EM)&=&{2\over 3}[\bar{u}_3\gamma_\mu u_3 +\sum_{i=1}^2(\bar{u}_i\gamma_\mu u_i + \bar{U}_i\gamma_\mu U_i \\ \nonumber
& & + \bar{U'}_i\gamma_\mu U'_i)] \\ \nonumber
& & -{1\over3}(\bar{d}_3\gamma_\mu d_3 + \bar{D}_3\gamma_\mu D_3 + \bar{D'}_3\gamma_\mu D'_3 \\ \nonumber
& & +\sum_{i=1}^2\bar{d}_i\gamma_\mu d_i) \\ \nonumber
& & -\sum_{\alpha=1}^3(\bar{e}^-_\alpha\gamma_\mu e^-_\alpha + \bar{E}^-_\alpha\gamma_\mu E^-_\alpha + \bar{E'}^-_\alpha\gamma_\mu E^{\prime -}_\alpha) \\ \label{emc}
&=&\sum_f q_f\bar{f}\gamma_\mu f,
\end{eqnarray}
\begin{eqnarray}
J_\mu(Z)&=&J_{\mu,L}(Z)-S^2_WJ_\mu(EM),\\ \label{lzpc}
J_\mu(Z')&=&J_{\mu,L}(Z')-T_WJ_\mu(EM),
\end{eqnarray}
\begin{eqnarray}\nonumber
J_\mu(Z'')&=&-\bar{D}_{3L}\gamma_\mu D_{3L} + \bar{D'}_{3L}\gamma_\mu D'_{3L} \\ \nonumber
& & +\sum_{i=1}^2(\bar{U}_{iL}\gamma_\mu 
U_{iL}-\bar{U'}_{iL}\gamma_\mu U'_{iL})\\ \label{lzppc}
& & -\sum_{\alpha=1}^3(\bar{E^-}_{\alpha L}\gamma_\mu E^-_{\alpha L}
-\bar{E}^{\prime -}_{\alpha L}\gamma_\mu E^{\prime -}_{\alpha L}),
\end{eqnarray}
\noindent where $e=gS_W=g_4S_W=g_XC_W\sqrt{1-T_W ^2/2}>0$ is the electric charge, 
$q_f$ is the electric charge of the fermion $f$ in units of $e$ and 
$J_\mu(EM)$ is the electromagnetic current. Notice that the $Z''_\mu$ 
current couples only to exotic fields. 

\begin{table*}
\caption{\label{tab3}The $Z_1^\mu\longrightarrow \bar{f}f$ couplings.}
\begin{ruledtabular}
\begin{tabular}{lcc}
$f$ & $g(f)_{1V}$ & $g(f)_{1A}$ \\ \hline
$u_3$ & $({1\over 2}-{4S_W^2 \over 3})(\cos\theta +
\Upsilon \sin\theta) $
& ${1\over 2} (\cos\theta + \Upsilon \sin\theta) $ \\ 
$d_3$ & $(-{1\over
2}+{2S_W^2\over 3})\cos\theta+{1\over 2}\Upsilon (C^2_W+\frac{S^2_W}{3})\sin\theta $ 
& $-{1\over 2}(\cos\theta - \Upsilon C_{2W}\sin\theta )$ \\
$D_3$ & ${2S_W^2\over 3}\cos\theta-{1\over 2}\Upsilon (1 
- {7 S_W^2 \over 3})\sin\theta $ &
$-{1\over 2}\Upsilon C_W^2\sin\theta $ \\  
$D'_3$ & ${2S_W^2\over 3}\cos\theta-{1\over 2}\Upsilon (1 
- {7 S_W^2 \over 3})\sin\theta $ &
$-{1\over 2}\Upsilon C_W^2\sin\theta $ \\
$d_{1,2}$ & $(-{1\over
2}+{2S_W^2\over 3})(\cos\theta +\Upsilon \sin\theta) $ & $-\frac{1}{2}(\cos\theta 
+ \Upsilon\sin\theta) $ \\
$u_{1,2}$& $({1\over 2}-{4S_W^2 \over 3})\cos\theta -
{1\over 2}\Upsilon (C^2_W +{5 S_W^2\over 3})\sin\theta $
& ${1\over 2} (\cos\theta - \Upsilon C_{2W}\sin\theta) $ \\ 
$U_{1,2}$& $-{4S_W^2\over 3}\cos\theta
+\frac{1}{2}\Upsilon (1-{11\over 3}S_W^2)\sin\theta $ &
$\frac{1}{2}\Upsilon C_W^2\sin\theta $ \\
$U'_{1,2}$& $-{4S_W^2\over 3}\cos\theta
+\frac{1}{2}\Upsilon (1-{11\over 3}S_W^2)\sin\theta $ &
$\frac{1}{2}\Upsilon C_W^2\sin\theta $ \\  
$\nu_{1,2,3}$& ${1\over 2}(\cos\theta
+\Upsilon\sin\theta)$ & $ {1\over 2}(\cos\theta
+\Upsilon\sin\theta)$\\
$e^-_{1,2,3}$ &
$(-{1\over 2}+2 S^2_W)\cos\theta+\Upsilon ({1\over 2}+ S^2_W)\sin\theta $ & $ -{1\over 2}(\cos\theta - \Upsilon C_{2W}\sin\theta )$ \\ 
$E^-_{1,2,3}$ & $2 S_W^2\cos\theta + \frac{1}{2}\Upsilon (-1+5 S^2_W)\sin\theta $ &
$-\frac{1}{2}\Upsilon C^2_W \sin\theta $ \\ 
$E^{\prime -}_{1,2,3}$ &
$2 S_W^2\cos\theta + \frac{1}{2}\Upsilon (-1+5 S^2_W)\sin\theta $ &
$-\frac{1}{2}\Upsilon C^2_W\sin\theta $ \\ 
\end{tabular}
\end{ruledtabular}
\end{table*}

\begin{table*}
\caption{\label{tab4}The $Z_2^\mu\longrightarrow \bar{f}f$ couplings.}
\begin{ruledtabular}
\begin{tabular}{lcc}
$f$ & $g(f)_{2V}$ & $g(f)_{2A}$ \\ \hline
$u_3$ & $-({1\over 2}-{4S_W^2 \over 3})(\sin\theta -
\Upsilon \cos\theta) $
& ${1\over 2} (-\sin\theta + \Upsilon \cos\theta) $ \\ 
$d_3$ & $({1\over 2}-{2S_W^2\over 3})\sin\theta+{1\over 2}\Upsilon (C^2_W+\frac{S^2_W}{3})\cos\theta $ 
& ${1\over 2}(\sin\theta + \Upsilon C_{2W}\cos\theta )$ \\
$D_3$ & $-{2S_W^2\over 3}\sin\theta-{1\over 2}\Upsilon (1 
- {7 S_W^2 \over 3})\cos\theta $ &
$-{1\over 2}\Upsilon C_W^2\cos\theta $ \\  
$D'_3$ & $-{2S_W^2\over 3}\sin\theta-{1\over 2}\Upsilon (1 
- {7 S_W^2 \over 3})\cos\theta $ &
$-{1\over 2}\Upsilon C_W^2\cos\theta $ \\
$d_{1,2}$ & $({1\over
2}-{2S_W^2\over 3})(\sin\theta -\Upsilon \cos\theta) $ & 
$-\frac{1}{2}(-\sin\theta 
+ \Upsilon\cos\theta) $ \\
$u_{1,2}$& $-({1\over 2}-{4S_W^2 \over 3})\sin\theta -
{1\over 2}\Upsilon (C^2_W +{5 S_W^2\over 3})\cos\theta $
& $-{1\over 2} (\sin\theta + \Upsilon C_{2W}\cos\theta) $ \\ 
$U_{1,2}$& ${4S_W^2\over 3}\sin\theta
+\frac{1}{2}\Upsilon (1-{11\over 3}S_W^2)\cos\theta $ &
$\frac{1}{2}\Upsilon C_W^2\cos\theta $ \\
$U'_{1,2}$& ${4S_W^2\over 3}\sin\theta
+\frac{1}{2}\Upsilon (1-{11\over 3}S_W^2)\cos\theta $ &
$\frac{1}{2}\Upsilon C_W^2\cos\theta $ \\  
$\nu_{1,2,3}$& ${1\over 2}(-\sin\theta
+\Upsilon\cos\theta)$ & $ {1\over 2}(-\sin\theta
+\Upsilon\cos\theta)$\\
$e^-_{1,2,3}$ &
$({1\over 2}-2 S^2_W)\sin\theta+\Upsilon ({1\over 2}+ S^2_W)\cos\theta $ & $ {1\over 2}(\sin\theta + \Upsilon C_{2W}\cos\theta )$ \\ 
$E^-_{1,2,3}$ & $-2 S_W^2\sin\theta + \frac{1}{2}\Upsilon (-1+5 S^2_W)\cos\theta $ &
$-\frac{1}{2}\Upsilon C^2_W \cos\theta $ \\ 
$E^{\prime -}_{1,2,3}$ &
$-2 S_W^2\sin\theta + \frac{1}{2}\Upsilon (-1+5 S^2_W)\cos\theta $ &
$-\frac{1}{2}\Upsilon C^2_W\cos\theta $ \\ 
\end{tabular}
\end{ruledtabular}
\end{table*}

The left-handed currents are
\begin{eqnarray} \nonumber
J_{\mu,L}(Z)&=&{1\over 2} [\bar{u}_{3L}\gamma_\mu u_{3L}-\bar{d}_{3L}\gamma_\mu d_{3L} \\ \nonumber
& & -\sum_{i=1}^2(\bar{d}_{iL}\gamma_\mu d_{iL}-\bar{u}_{iL}\gamma_\mu u_{iL})\\ \nonumber
& & +\sum_{\alpha=1}^3(\bar{\nu}_{e\alpha L}\gamma_\mu \nu_{e\alpha L} - \bar{e}^-_{\alpha L}\gamma_\mu e^-_{\alpha L})] \\ \label{lhz} 
&=&\sum_f T_{4f}\bar{f}_L\gamma_\mu f_L ,
\end{eqnarray}

\begin{eqnarray}\nonumber
J_{\mu,L}(Z')&=&(2T_{W})^{-1}\lbrace(1+T^2_W)\bar{u}_{3L}\gamma_\mu u_{3L} \\ \nonumber 
& & +(1-T^2_W)\bar{d}_{3L}\gamma_\mu d_{3L} \\ \nonumber
& & -\bar{D}_{3L}\gamma_\mu D_{3L}-\bar{D'}_{3L}\gamma_\mu D'_{3L} \\ \nonumber
& & -\sum_{i=1}^2\lbrack (1+T^2_W)\bar{d}_{iL}\gamma_\mu d_{iL} \\ \nonumber
& & +(1-T^2_W)\bar{u}_{iL}\gamma_\mu u_{iL} \\ \nonumber
& & -\bar{U}_{iL}\gamma_\mu U_{iL}-\bar{U'}_{iL}\gamma_\mu U'_{iL}\rbrack \\ \nonumber 
& & +\sum_{\alpha=1}^3\lbrack (1+T^2_W)\bar{\nu}_{\alpha L}\gamma_\mu \nu_{\alpha L} \\ \nonumber
& & +(1-T^2_W)\bar{e}^-_{\alpha L}\gamma_\mu e^-_{\alpha L} \\ \nonumber
& & -\bar{E}^-_{\alpha L}\gamma_\mu E^-_{\alpha L} -\bar{E}^{\prime -}_{\alpha L}\gamma_\mu E^{\prime -}_{\alpha L}\rbrack \rbrace
\\ \label{lhzp}
&=&\sum_f T'_{4f}\bar{f}_L\gamma_\mu f_L,
\end{eqnarray}
where $T_{4f}=Dg(1/2,-1/2,0,0)$ is the third component of the weak isospin
and $T'_{4f}=(1/2T_W)Dg(1+T^2_W, 1-T^2_W, -1, -1)$= $T_W\lambda_3/2 +(1/T_W)(\lambda_8/\sqrt{3}+\lambda_{15}/\sqrt{6})$ is a convenient $4\times 4$ diagonal matrix, acting both of them on the representation 4 of $SU(4)_L$. The current $J_\mu(Z)$ is clearly recognized as the generalization of the neutral current of the SM. Thus, we identify $Z_\mu$ as the neutral gauge boson of the SM, which is consistent with Eqs.~(\ref{fzzp}) and (\ref{y}).

From Eq.~(\ref{lhzp}) we can see that the left-handed couplings of $Z^\prime$ to the third family of quarks are different from the ones to the first two families. This induces FCNC at tree level transmitted by the $Z^\prime$ boson.

The couplings between the mass eigenstates $Z_1^\mu$, $Z_2^\mu$ and the fermion fields are obtained from
\begin{eqnarray} \nonumber
-{\cal L}^{NC}_{Z_1,Z_2}&=&\frac{g_4}{2C_W}\sum_{i=1}^2Z_i^\mu\sum_f\{\bar{f}\gamma_\mu
[a_{iL}(f)(1-\gamma_5)\\ \nonumber & & 
+a_{iR}(f)(1+\gamma_5)]f\} \\ \nonumber
      &=&\frac{g_4}{2C_W}\sum_{i=1}^2Z_i^\mu\sum_f\{\bar{f}\gamma_\mu
      [g(f)_{iV}-g(f)_{iA}\gamma_5]f\},
\end{eqnarray}
\noindent
where
\begin{eqnarray} \nonumber
a_{1L}(f)&=&\cos\theta(T_{4f}-q_fS^2_W)\\ \nonumber & &
+\frac{g_X\sin\theta
C_W}{g_4\sqrt{2}} (T'_{4f}-q_fT_W)\;, \\ \nonumber
a_{1R}(f)&=&-q_fS_W\left(\cos\theta
S_W+\frac{g_X\sin\theta}{g_4\sqrt{2}}\right)\;,\\ \nonumber
a_{2L}(f)&=&-\sin\theta(T_{4f}-q_fS^2_W)\\ \nonumber & &
+\frac{g_X\cos\theta
C_W}{g_4\sqrt{2}} (T'_{4f}-q_fT_W)\;, \\ \label{a}
a_{2R}(f)&=&q_fS_W\left(\sin\theta S_W-\frac{g_X\cos\theta}{g_4\sqrt{2}}\right),
\end{eqnarray}
and
\begin{eqnarray} \nonumber
g(f)_{1V}&=&\cos\theta(T_{4f}-2S_W^2q_f)\\ \nonumber & &
+\frac{g_X\sin\theta}{g_4\sqrt{2}}
(T'_{4f}C_W-2q_fS_W)\;, \\ \nonumber
g(f)_{2V}&=&-\sin\theta(T_{4f}-2S_W^2q_f)\\ \nonumber & &
+\frac{g_X\cos\theta}{g_4\sqrt{2}}
(T'_{4f}C_W-2q_fS_W) \;,\\ \nonumber g(f)_{1A}&=&\cos\theta
T_{4f}+\frac{g_X\sin\theta}{g_4\sqrt{2}}T'_{4f}C_W\;, \\ \label{g}
g(f)_{2A}&=&-\sin\theta
T_{4f}+\frac{g_X\cos\theta}{g_4\sqrt{2}}T'_{4f}C_W.
\end{eqnarray}
The values of $g_{iV},\; g_{iA}$ with $i=1,2$ are listed in Tables \ref{tab3} and \ref{tab4}, where $\Upsilon=1/\sqrt{2-3S^2_W}$ and $C_{2W} = C^2_W - S^2_W$.

Note that in the limit $\theta\rightarrow 0$ the couplings of $Z_1^\mu$ to 
the ordinary leptons and quarks coincide with the ones in the SM. This allows us to test the new physics beyond the SM predicted by the model.

\section{\label{sec:sec4}Fermion spectrum}
In order to avoid mixing between ordinary and exotic fermions and, at the same time, to generate a consistent mass spectrum, we follow the same strategy than in Ref.~\cite{spp} (see also Ref.~\cite{su42}) and combine the set of Higgs scalars introduced in Sec.~\ref{sec:sec2} with an anomaly-free discrete $Z_2$ symmetry \cite{KW}, with the following assignments of $Z_2$ charge $q_Z$
\begin{eqnarray} \nonumber
q_Z(Q_{\alpha L}, u^c_{\alpha L}, d^c_{\alpha L}, L_{\alpha L}, e^c_{\alpha L}, \phi_1, \phi_2)&=& 0, \\ \label{z2d}
q_Z(U^c_{iL}, U^{\prime c}_{iL}, D^c_{3L}, D^{\prime c}_{3L}, E^c_{\alpha L}, E^{\prime c}_{\alpha L}, \phi_3, \phi_4)&=& 1,
\end{eqnarray}
where $\alpha = 1,2,3$ and $i=1,2$. Notice that ordinary fermions are not affected by this discrete symmetry.

The gauge invariance and the $Z_2$ symmetry allow for the following Yukawa lagrangians for quarks and for charged leptons, respectively:
 
\begin{eqnarray} \nonumber
{\cal L}^Q_Y&=& Q^T_{3L}C[\phi_1 \sum_{\alpha=1}^3 
h^u_{3\alpha}u^c_{\alpha L} + \phi_2 \sum_{\alpha=1}^3 
h^d_{3\alpha}d^c_{\alpha L} \\ \nonumber
&+& \sum_{k=3}^4\phi_k(h^{(k)D}_{33}D^c_{3L}+h^{(k)D'}_{33}D^{\prime c}_{3L})] \\ \nonumber
&+& \sum_{i=1}^2Q^T_{iL}C[\phi^*_1 \sum_{\alpha=1}^3h^d_{i\alpha} d^c_{\alpha L} + \phi^*_2\sum_{\alpha=1}^3h^u_{i\alpha}u^c_{\alpha L} \\ \nonumber
&+& \sum_{k=3}^4\phi^*_k\sum_{j=1}^2(h^{(k)U}_{ij}U^c_{jL}+h^{(k)U'}_{ij}U^{\prime c}_{jL})] 
+ h.c., \\ \label{mq}
\end{eqnarray}

\begin{eqnarray}\nonumber
{\cal L}_Y^l&=&\sum_{\alpha=1}^3\sum_{\beta=1}^3 
L_{\alpha L}^TC[\phi_2 h^e_{\alpha\beta}e^+_{\beta L} \\ \nonumber
& & +\sum_{k=3}^4\phi_k(h^{(k)E}_{\alpha\beta}E^+_{\beta L}+h^{(k)E'}_{\alpha\beta}E^{\prime +}_{\beta L})] + h.c., \\ \label{ml}
\end{eqnarray}
where the $h's$ are Yukawa couplings and $C$ is the charge conjugate 
operator.

The Lagrangian ${\cal L}^Q_Y$ produces for up- and down-type quarks, in the basis $(u_1,u_2,u_3,U_1,U_2,U'_1,U'_2)$ and $(d_1,d_2,d_3,D_3,D'_3)$ respectively, block diagonal mass matrices of the form
\begin{widetext}
\begin{eqnarray}\label{mumd}
M_{uU}=\left(\begin{array}{cc}
M^u_{3\times 3} &  0\\
0 &  M^U_{4\times 4}\end{array}\right) \qquad \mbox{and} & \qquad
M_{dD}=\left(\begin{array}{cc}
M^d_{3\times 3} &  0\\
0 &  M^D_{2\times 2}\end{array}\right),
\end{eqnarray}
where
\begin{eqnarray}\label{up}
M^u=\left(\begin{array}{ccc}
h^u_{11}v^\prime & h^u_{12}v^\prime & h^u_{13}v^\prime \\
h^u_{21}v^\prime & h^u_{22}v^\prime & h^u_{23}v^\prime \\
h^u_{31}v & h^u_{32}v & h^u_{33}v \end{array}\right) \qquad \mbox{and} & \qquad
M^U=\left(\begin{array}{cccc}
h^{(3)U}_{11}V & h^{(3)U}_{12}V & h^{(3)U'}_{11}V & h^{(3)U'}_{12}V\\
h^{(3)U}_{21}V & h^{(3)U}_{22}V & h^{(3)U'}_{21}V & h^{(3)U'}_{22}V\\
h^{(4)U}_{11}V^\prime & h^{(4)U}_{12}V^\prime & h^{(4)U'}_{11}V^\prime & h^{(4)U'}_{12}V^\prime\\
h^{(4)U}_{21}V^\prime & h^{(4)U}_{22}V^\prime & h^{(4)U'}_{21}V^\prime & h^{(4)U'}_{22}V^\prime
\end{array}\right);
\end{eqnarray}

\begin{eqnarray}\label{down}
M^d=\left(\begin{array}{ccc}
h^d_{11}v & h^d_{12}v & h^d_{13}v \\
h^d_{21}v & h^d_{22}v & h^d_{23}v \\
h^d_{31}v^\prime & h^d_{32}v^\prime & h^d_{33}v^\prime \end{array}\right) \qquad \mbox{and} & \qquad
M^D=\left(\begin{array}{cc}
h^{(3)D}_{33}V & h^{(3)D'}_{33}V\\
h^{(4)D}_{33}V^\prime & h^{(4)D'}_{33}V^\prime\end{array}\right).
\end{eqnarray}
\end{widetext}
For the charged leptons the Lagrangian ${\cal L}_Y^l$, in the basis $(e_1,e_2,e_3,E_1,E_2,E_3,E'_1,E'_2,E'_3)$, also produces a block diagonal mass matrix 
\begin{equation}\label{mlep}
M_{eE}=\left(\begin{array}{cc}
M^e_{3\times 3} &  0\\
0 &  M^E_{6\times 6}\end{array}\right),
\end{equation}
where the submatrices are
\begin{equation}
M^e_{\alpha \beta} = h^e_{\alpha \beta} v^\prime,
\end{equation}
and
\begin{widetext}
\begin{equation}\nonumber
M^{E} = \left(\begin{array}{cccccc}
h^{(3)E}_{11}V & h^{(3)E}_{12}V & h^{(3)E}_{13}V & h^{(3)E'}_{11}V & h^{(3)E'}_{12}V & h^{(3)E'}_{13}V\\
h^{(3)E}_{21}V & h^{(3)E}_{22}V & h^{(3)E}_{23}V & h^{(3)E'}_{21}V & h^{(3)E'}_{22}V & h^{(3)E'}_{23}V\\
h^{(3)E}_{31}V & h^{(3)E}_{32}V & h^{(3)E}_{33}V & h^{(3)E'}_{31}V & h^{(3)E'}_{32}V & h^{(3)E'}_{33}V\\
h^{(4)E}_{11}V^{\prime} & h^{(4)E}_{12}V^{\prime} & h^{(4)E}_{13}V^{\prime} & h^{(4)E'}_{11}V^{\prime} & h^{(4)E'}_{12}V^{\prime} & h^{(4)E'}_{13}V^{\prime}\\
h^{(4)E}_{21}V^{\prime} & h^{(4)E}_{22}V^{\prime} & h^{(4)E}_{23}V^{\prime} & h^{(4)E'}_{21}V^{\prime} & h^{(4)E'}_{22}V^{\prime} & h^{(4)E'}_{23}V^{\prime}\\
h^{(4)E}_{31}V^{\prime} & h^{(4)E}_{32}V^{\prime} & h^{(4)E}_{33}V^{\prime} & h^{(4)E'}_{31}V^{\prime} & h^{(4)E'}_{32}V^{\prime} & h^{(4)E'}_{33}V^{\prime}
\end{array}\right)
\end{equation}
\end{widetext}
The mass matrices in Eqs.~(\ref{mumd}) and (\ref{mlep}) exhibit a simple mass splitting between ordinary and exotic charged fermions and show that all the charged fermions in the model acquire masses at the three level. By a judicious tuning of the Yukawa couplings and of the mass scales $v$ and $v^\prime$, a consistent mass spectrum in the ordinary charged sector can be obtained. All the exotic charged fermions acquire masses at the scale $V \sim V^\prime \gg v_{EW} = 174$~GeV.

Notice that the tensor product form of the mass matrices $M_{uU}$ and $M_{dD}$ in Eq.~(\ref{mumd}), implies that they are diagonalized by unitary matrices which are themselves tensor products of unitary matrices. So, our discrete $Z_2$ symmetry has the important bonus of avoiding violation of unitarity of the CKM mixing matrix.

The neutral leptons $\nu^0_{e\alpha}$, $\alpha=1,2,3$, remain massless. However, neutrino masses and mixing can be implemented in the model by introducing one or more Weyl singlet states $N^0_{L,n}\sim [1,1,0],\; n=1,2,...$, without violating our assumptions, neither the anomaly constraint relations, because singlets with no $X$-charges are as good as not being present as far as anomaly cancellation is concerned.

\section{\label{sec:sec5}Constraints on the parameters of the model}

\subsection{\label{sec:sec5a}Constrains on the $Z^\mu-Z'^{\mu}$ mixing angle $\theta$ and the $Z^{\mu}_2$ mass} 
In this section, using measurements at the Z-pole and APV data, we shall set bounds on the mass of the new neutral gauge boson $Z^{\mu}_2$, and its mixing angle $\theta$ with the ordinary neutral gauge boson. Next, these bounds will be compared with the ones resulting from FCNC effects present in the model.

\subsubsection{\label{sec:sec5a1}Bounds from Z-pole observables and APV data} 
Let us start constrainig the parameter space ($\theta-M_{Z_2}$) by using observables measured at the $Z$-pole from CERN $e^+e^-$ collider (LEP), SLAC Linear Collider (SLC), and atomic parity violation constraints which are given in Table \ref{tab5}. 

The partial decay width for $Z^{\mu}_1\rightarrow f\bar{f}$ is given by
 
\begin{eqnarray}\nonumber
\Gamma(Z^{\mu}_1\rightarrow f\bar{f})&=&\frac{N_C G_F
M_{Z_1}^3}{6\pi\sqrt{2}}\rho \Big\{\frac{3\beta-\beta^3}{2}
[g(f)_{1V}]^2 \\ \nonumber
& + & \beta^3[g(f)_{1A}]^2 \Big\}(1+\delta_f)R_{EW}R_{QCD}, \\ \label{ancho}
\end{eqnarray}

\noindent 
where $f$ is an ordinary SM fermion, $Z^\mu_1$ is the physical gauge boson
observed at LEP, $N_C=1$ for leptons while for quarks
$N_C=3(1+\alpha_s/\pi + 1.405\alpha_s^2/\pi^2 - 12.77\alpha_s^3/\pi^3)$,
where the 3 is due to color and the factor in parentheses represents the
universal part of the QCD corrections for massless quarks 
(for fermion mass effects and further QCD corrections which are 
different for vector and axial-vector partial widths, see 
Ref.~\cite{kuhn}); $R_{EW}$ are the electroweak corrections which include the leading order QED corrections given by $R_{QED}=1+3\alpha/(4\pi)$. $R_{QCD}$ are further QCD corrections (for a comprehensive review see Ref.~\cite{leike} and references therein), and $\beta=\sqrt{1-4 m_f^2/M_{Z_1}^2}$ is a kinematic factor which can be taken equal to $1$ for all the SM fermions except for the bottom quark. 
The factor $\delta_f$ contains the one loop vertex
contribution which is negligible for all fermion fields except for the 
bottom quark for which the contribution coming from the top quark at the 
one loop vertex radiative correction is parametrized as $\delta_b\approx 
10^{-2} [-m_t^2/(2 M_{Z_1}^2)+1/5]$ \cite{pich}. The $\rho$ parameter 
can be expanded as $\rho = 1+\delta\rho_0 + \delta\rho_V$ where the 
oblique correction $\delta\rho_0$ is given by
$\delta\rho_0\approx 3G_F m_t^2/(8\pi^2\sqrt{2})$, and $\delta\rho_V$ is 
the tree level contribution due to the $(Z_{\mu} - Z'_{\mu})$ mixing which 
can be parametrized as $\delta\rho_V\approx
(M_{Z_2}^2/M_{Z_1}^2-1)\sin^2\theta$. Finally, $g(f)_{1V}$ and $g(f)_{1A}$
are the coupling constants of the physical $Z_1^\mu$ field with ordinary
fermions which are listed in Table \ref{tab3}.

The ratios of partial widths are defined as $R_l\equiv \Gamma_Z(\mbox{had})/\Gamma(l^+l^-)$ for $l=e,\mu,\tau$, and $R_\eta\equiv \Gamma_\eta/\Gamma_Z(\mbox{had})$ for $\eta=b,c$. We shall use the experimental values \cite{pdg}:
$\alpha_s(M_Z)=0.1198$, $\alpha(M_Z)^{-1}=127.918$, and $\sin^2\theta_W=0.2312$. 

As a first result notice from Table \ref{tab3} that, due to universality in the lepton sector, our model predicts $R_e=R_\mu=R_\tau$, in agreement with the experimental results in Table \ref{tab5}. 

\begin{table}
\caption{\label{tab5}Experimental data and SM values for some observables related to neutral currents.}
\begin{ruledtabular}
\begin{tabular}{lcl}
& Experimental results & SM \\ \hline
$\Gamma_Z$(GeV)  & $2.4952 \pm 0.0023$  &  $2.4968 \pm 0.0011$  \\   
$\Gamma(had)$ (GeV)  & $1.7444 \pm 0.0020$ & $1.7434 \pm 0.0010$ \\ 
$\Gamma(l^+l^-)$ (MeV) & $83.984\pm 0.086$ & $83.996 \pm 0.021$ \\
$R_e$ & $20.804\pm 0.050$ & $20.756\pm 0.011$ \\ 
$R_\mu$ & $20.785\pm 0.033$ & $20.756\pm 0.011$ \\ 
$R_\tau$ & $20.764\pm 0.045$ & $20.801\pm 0.011$ \\ 
$R_b$ & $0.21629\pm 0.00066$ & $0.21578\pm 0.00010$ \\ 
$R_c$ & $0.1721\pm 0.0030$ & $0.17230\pm 0.00004$ \\ 
$Q_W^{Cs}$ & $-72.62\pm 0.46$ & $-73.17\pm 0.03$ \\
$M_{Z_{1}}$(GeV) & $ 91.1876 \pm 0.0021 $ & $ 91.1874 \pm 0.0021 $ \\
$m_t$(GeV) & $172.7 \pm 2.9 \pm 0.6$ & $ 172.7 \pm 2.8 $ \\
\end{tabular}
\end{ruledtabular}
\end{table}
The effective weak charge in atomic parity violation, $Q_W$, can be 
expressed as 

\begin{equation}
Q_W=-2\left[(2Z+N)c_{1u}+(Z+2N)c_{1d}\right], 
\end{equation}
\noindent
where $Z$ is the number of protons and $N$ the number of neutrons in the atomic nucleus, and $c_{1q}=2g(e)_{1A}g(q)_{1V}$. The theoretical value for $Q_W$ for the Cesium atom is given by \cite{guena} $Q_W(^{133}_{55}Cs)=-73.17\pm0.03 + \Delta Q_W$, where the contribution of new physics is included in $\Delta Q_W$ and can be written as \cite{durkin}

\begin{equation}\label{DQ} 
\Delta 
Q_W=\left[\left(1+4\frac{S^4_W}{1-2S^2_W}\right)Z-N\right]\delta\rho_V
+\Delta Q^\prime_W,
\end{equation}

The term $\Delta Q^\prime_W$ is model dependent. In particular, is mostly a function of the couplings $g(q)_{2V}$ and $g(q)_{2A}$ ($q=u,d$) of the first family of quarks to the new neutral gauge boson $Z_2$ \cite{alta}. So, the new physics in $\Delta Q^\prime_W$ depends on which family of quarks transform differently under the gauge group. Taking the third generation being different and using $g(e)_{iA}$ and $g(q)_{iV}$, $i=1,2$ from Tables \ref{tab3} and \ref{tab4}, we obtain
\begin{equation}
\Delta Q_W^\prime=(10.35 Z + 10.76 N) \sin\theta + (4.94 Z + 4.18 N)
\frac{M^2_{Z_1}}{M^2_{Z_2}}\; .
\end{equation}

The discrepancy between the SM and the experimental data for $\Delta Q_W$ 
is given by \cite{guena}

\begin{equation}
\Delta Q_W=Q^{exp}_W-Q^{SM}_W=0.45\pm 0.48,
\end{equation}
which is $1.1\; \sigma$ away from the SM predictions.

Introducing the expressions for $Z$-pole observables in Eq.~(\ref{ancho}), 
with $\Delta Q_W$ in terms of new physics in Eq.(\ref{DQ}) and using 
experimental data from LEP, SLC and atomic parity violation (see Table 
\ref{tab5}), we do a $\chi^2$ fit and find the best allowed region in the $(\theta-M_{Z_2})$ plane at $95\%$ confidence level (C.L.). In 
Fig.~1 we display this region which gives us the constraints 
\begin{equation} 
-0.00017 \leq \theta \leq 0.0007, \;\;\; 2.02\; {\mbox TeV} \leq M_{Z_2}. \label{bounds}
\end{equation} 
This shows that the mass of the new neutral gauge boson is compatible 
with the bound obtained in $p\bar{p}$ collisions at the Fermilab 
Tevatron \cite{abe}. In the limit $\vert \theta \vert \rightarrow 0$, $M_{Z_2}$ peaks at a finite value larger than $100$ TeV which still copes with experimental constraints on the $\rho$ parameter.

As already mentioned, the bounds in Eq.~(\ref{bounds}) depend on the fact that it is the third generation of quarks the one transforming differently under $SU(4)_L\otimes U(1)_X$. We have checked that a different choice raises the lower bound on $M_{Z_2}$ in Eq.~(\ref{bounds}) to a value larger than 3 TeV. So, the third family must be different in order to keep this 
lower bound as low as possible.

\subsubsection{\label{sec:sec5a2}Bounds from FCNC processes} 
The discrete $Z_2$ symmetry introduced in Sec.~\ref{sec:sec3} not only produces a simple mass splitting between ordinary and exotic fermions, but also avoids unitarity violation of the CKM matrix arising from their mixing. Because the right-handed quarks transform as $SU(4)_L$ singlets, they are generation universal and, consequently, they couple diagonally to the neutral gauge bosons $Z$ and $Z^\prime$. However, there is additional source of FCNC due to the fact that anomaly cancellation among the families forces us to have one family of left-handed quarks in the weak basis, transforming differently from the other two. Moreover, since each flavor couples to more than one Higgs 4-plet, FCNC coming from the scalar sector are also present. Because this last contribution depends on a large number of arbitrary parameters, is not very useful in order to constraint the model and we will ignore it.

Regarding the left-handed interaction of quarks, Eq.~(\ref{lhz}) shows that the couplings to the $Z$ boson remain flavor conserving. Then, neglecting the scalar contribution and because of the $Z_2$ symmetry, the only source of FCNC in the model is in the left-handed interactions of ordinary quarks with the new neutral gauge boson $Z^\prime$. For their study we will follow the analysis presented in Refs.~\cite{liu,jars} where bounds coming from neutral meson mixing are obtained in the context of the so-called ``minimal 3-3-1 model".

Using the expression for the $Y$ hypercharge of the SM: $Y/2=T_{8L}/\sqrt{3}+T_{15L}/\sqrt{6}+X$, the couplings of $Z^\prime$ to left-handed quarks in Eqs.~(\ref{lzpc}) and (\ref{lhzp}), can be written in a more convenient fashion for 4-plets as

\begin{equation}\label{Lzprime}
{\cal L}(Z^\prime)=\frac{g}{2C_W} \frac{1}{\sqrt{6}\sqrt{2-3S^2_W}}Z^\prime_\mu J^\mu(Z^\prime),
\end{equation}
with

\begin{equation}
J^\mu(Z^\prime)=\sum_f \bar{f}\gamma^\mu[\sqrt{6}S^2_WY-4C^2_WT_L]P_Lf.
\end{equation}
where $P_L$ is the left-handed projector and $T_L=\sqrt{2}T_{8L}+T_{15L}$.

The value of the operator $T_L$ is not the same for 4-plets than for $4^*$-plets. Then, the flavor changing interaction can be written, for ordinary up- and down-type quarks in the weak basis, as
\begin{equation}\label{zfc}
J^\mu(Z^\prime)_{FCNC}=-4C^2_W\sum_{q^\prime} \bar{q^\prime}\gamma^\mu[T_L(4)-T_L(4^*)]P_Lq^\prime.
\end{equation}

Using (\ref{Lzprime}) and (\ref{zfc}) we have
\begin{eqnarray}\nonumber
{\cal L}(Z^\prime)_{FCNC}&=&-\frac{gC^2_W}{\sqrt{2-3S^2_W}}(\sin\theta Z^\mu_1+\cos\theta Z^\mu_2) \\ \label{Lfcnc}
& & \times\sum_{q^\prime} \bar{q^\prime}\gamma_\mu P_L q^\prime,
\end{eqnarray}
where, by using Eq.~(\ref{mixing}), we have included the mass eigenstates $Z_1$ and $Z_2$. 

\begin{table}
\caption{\label{tab6}Values of the experimental and theoretical quantities used as input parameters for FCNC processes.}
\begin{ruledtabular}
\begin{tabular}{llc}
& Value & Reference \\ \hline
$\Delta m_K$ (GeV) & $3.483(6)\cdot 10^{-15}$ & \cite{pdg} \\   
$m_{K^0}$ (MeV) & $497.65(2)$ & \cite{pdg} \\ 
$f_K\sqrt{B_K}$ (MeV) & $143(7)$ & \cite{hasi} \\ \hline
$\Delta m_{B_d}$ ($\mbox{ps}^{-1}$) & $0.508(4)$ & \cite{pdg} \\ 
$m_{B_d}$ (GeV) & $5.2794(5)$ & \cite{pdg} \\ 
$f_{B_d}\sqrt{B_{B_d}}$ (MeV) & $214(38)$ & \cite{hasi} \\ \hline
$\Delta m_{B_s}$ ($\mbox{ps}^{-1}$) & $17.77(12)$ & \cite{abuaba} \\ 
$m_{B_s}$ (GeV) & $5.370(2)$ & \cite{pdg} \\ 
$f_{B_s}\sqrt{B_{B_s}}$ (MeV) & $262(35)$ & \cite{hasi} \\ \hline
$\Delta m_D$($\mbox{ps}^{-1}$) & $11.7(6.8)\cdot 10^{-3}$ & \cite{ciu} \\   
$m_{D^0}$ (GeV) & $1.8645(4)$ & \cite{pdg} \\
$f_D\sqrt{B_D}$ (MeV) & $241(24)$ & \cite{arlin} \\ \hline
$m_u(M_Z)$ (MeV) & $2.33^{+0.42}_{-0.45}$ & $$ \\   
$m_c(M_Z)$ (MeV) & $677^{+56}_{-61}$ & $$ \\ 
$m_t(M_Z)$ (GeV) & $181\pm 13$ & $$ \\
$m_d(M_Z)$ (MeV) & $4.69^{+0.60}_{-0.66}$ & $$ \\
$m_s(M_Z)$ (MeV) & $93.4^{+11.8}_{-13.0}$ & $$ \\
$m_b(M_Z)$ (GeV) & $3.00\pm 0.11$ & \cite{fusa} \\
\end{tabular}
\end{ruledtabular}
\end{table}

In order to consider constraints coming from experimental data in the $K^0-\bar{K}^0$, $B^0_d-\bar{B}^0_d$, $B^0_s-\bar{B}^0_s$ and $D^0-\bar{D}^0$ systems, we first notice that the mass matrices $M^u$ and $M^d$ in Eqs.~(\ref{up}) and (\ref{down}) are diagonalized by biuntary transformations $U_{L,R}$ and $V_{L,R}$, respectively, with $V_{CKM}= U^{\dagger}_LV_L$ being the CKM mixing matriz. Then, in terms of mass eigenstates, Eq.~(\ref{Lfcnc}) produces the following effective Hamiltonian for the tree-level neutral meson mixing interactions 
\begin{eqnarray}\nonumber
{\cal H}^{(\alpha,\beta)}_{eff}&=&\frac{2\sqrt{2}G_F C^4_W \cos^2\theta}{2-3S^2_W}(V^*_{L3\alpha}V_{L3\beta})^2 \\ \label{heff}
& & \times \biggl(\frac{M^2_{Z_1}}{M^2_{Z_2}}+\tan^2\theta\biggr)[\bar{\alpha}\gamma_\mu P_L\beta]^2,
\end{eqnarray}
where $(\alpha,\beta)$ must be replaced by $(d,s)$, $(d,b)$, $(s,b)$ and $(u,c)$ for the $K^0-\bar{K}^0$, $B^0_d-\bar{B}^0_d$, $B^0_s-\bar{B}^0_s$ and $D^0-\bar{D}^0$ systems, respectively, and $V_L$ must be replaced by $U_L$ for the neutral $D^0-\bar{D}^0$ system. The family index 3 in $V_L$ makes explicit that we have singled out the third family of quarks as the one transforming differently.

The effective Hamiltonian gives the following contribution to the mass difference $\Delta m_K$ 
\begin{eqnarray}\nonumber
\frac{\Delta m_K}{m_K}&=&\frac{4\sqrt{2}G_F C^4_W \cos^2\theta}{3(2-3S^2_W)}\mbox{Re} [(V^*_{L3d}V_{L3s})^2] \\ \label{mdifK}
& & \times \;\eta_K \biggl(\frac{M^2_{Z_1}}{M^2_{Z_2}}+\tan^2\theta\biggr)B_Kf^2_K,
\end{eqnarray}
while for the $B^0_d-\bar{B}^0_d$, $B^0_s-\bar{B}^0_s$ and $D^0-\bar{D}^0$ systems, we have
\begin{eqnarray}\nonumber
\frac{\Delta m_B}{m_B}&=&\frac{4\sqrt{2}G_F C^4_W \cos^2\theta}{3(2-3S^2_W)}\vert V^*_{L3\alpha}V_{L3\beta}\vert^2 \\ \label{mdifB}
& & \times \;\eta_B \biggl(\frac{M^2_{Z_1}}{M^2_{Z_2}}+\tan^2\theta\biggr)B_Bf^2_B,
\end{eqnarray}

\begin{eqnarray}\nonumber
\frac{\Delta m_D}{m_D}&=&\frac{4\sqrt{2}G_F C^4_W \cos^2\theta}{3(2-3S^2_W)}\vert U^*_{L3u}U_{L3c}\vert^2 \\ \label{mdifD}
& & \times \;\eta_D \biggl(\frac{M^2_{Z_1}}{M^2_{Z_2}}+\tan^2\theta\biggr)B_Df^2_D,
\end{eqnarray}
\noindent
where $B$ stands for $B_d$ or $B_s$. $B_m$ and $f_m$ ($m=K, B_d, B_s, D$) are the bag parameter and decay constant of the corresponding neutral meson. The $\eta$'s are QCD corrections which, at leading order, can be taken equal to the ones of the SM \cite{blanke}, that is: $\eta_K\simeq \eta_D\simeq 0.57$, $\eta_{B_d} = \eta_{B_s}\simeq 0.55$ \cite{gilman}. 

Several sources, different from the tree-level $Z_2$ exchange, may contribute to the mass differences and it is not possible to disentangle the $Z_2$ contribution from other effects. Due to this, several authors consider reasonable to assume that the $Z_2$ exchange contribution must not be larger than the experimental values \cite{liu}. In this work, instead, we apply the more conservative criterion of Ref.~\cite{blanke} by using as deviations from the central value: $\pm 50\%$ for $\Delta m_K$, and $\pm 40\%$ for $\Delta m_{B_d}$ and $\Delta m_{B_s}$. We then assume that the $Z_2$ exchange contribution must not be larger than these deviations. In the $D$ system, the SM short distance contribution to $\Delta m_D$ is very small compared to the experimental value. Thus, in this case we only require that the $Z_2$ contribution must not exceed the observed value.

Since the complex numbers $V_{Lij}$ and $U_{Lij}$ can not be estimated from the present experimental data, and in order to compare with the bounds in Eq.~(\ref{bounds}), we assume the Fritzsch ansatz for the quark mass matrices \cite{frit}, which implies (for $i\leq j$) $V_{Lij}=\sqrt{m_i/m_j}$, and similary for $U_L$ \cite{cheng} (CP violating phases in the mixing matrices will not be considered here). 
To obtain bounds on $M_{Z_2}$ from Eqs.~(\ref{mdifK}), (\ref{mdifB}) and (\ref{mdifD}), we use updated experimental and theoretical values for the input parameters as shown in Table~\ref{tab6}, where the quark masses are given at Z-pole. 

\begin{figure} 
\begin{center}
\resizebox{0.98\textwidth}{!}{
\includegraphics{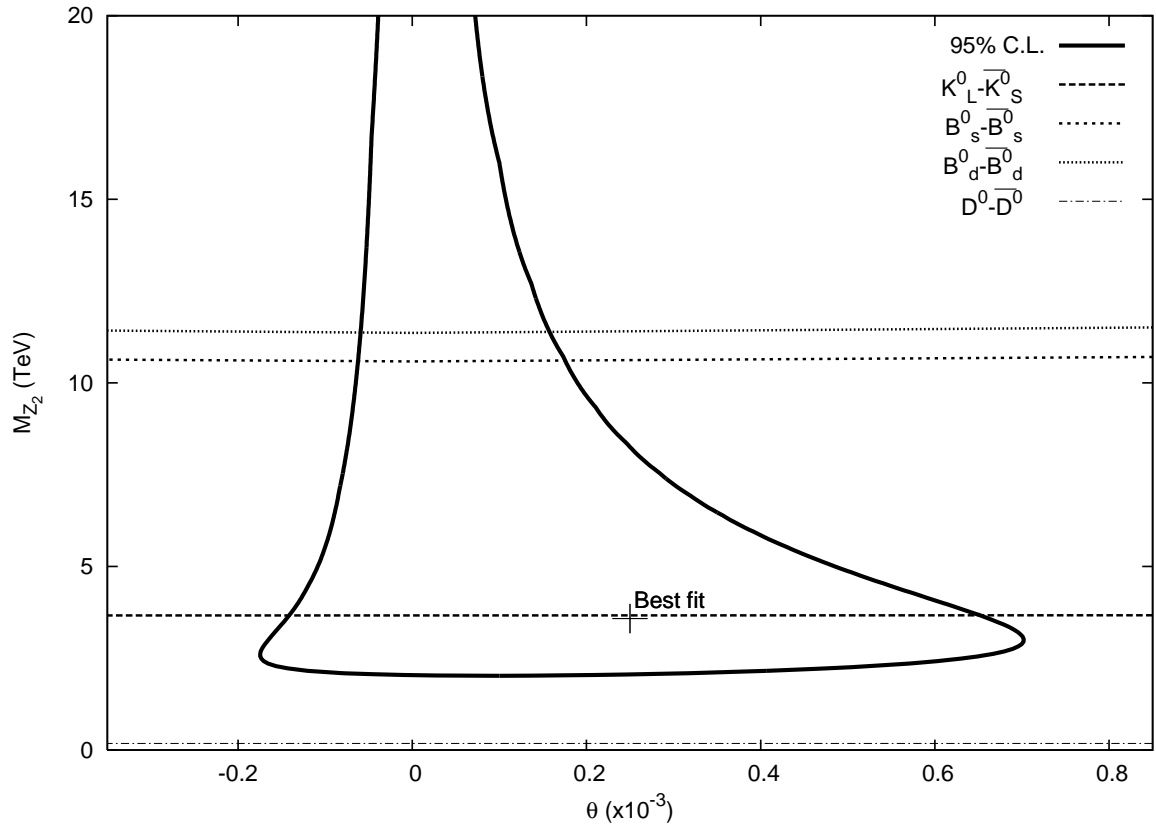}
}
\caption{\label{fig1} Contour plot displaying the allowed region for
$\theta$ vs. $M_{Z_2}$ at $95\%$ C.L. from LEP, SLC and APV data. The horizontal lines delimit from below the regions of the parameter space allowed by FCNC constraints.}
\end{center}
\end{figure}

In Fig.~\ref{fig1} are also shown the contours delimiting from below the regions in the parameter space ($\theta$-$M_{Z_2}$) where the $Z_2$ contribution to the mass differences satisfies the conditions mentioned above. These contours determine lower bounds on the $Z_2$ mass. The results are:
\begin{eqnarray}\nonumber
K^0-\bar{K}^0: & M_{Z_2}>3.66\; \mbox{TeV}, \\ \nonumber
B^0_d-\bar{B}^0_d: & M_{Z_2}>11.54 \;\mbox{TeV}, \\ \nonumber
B^0_s-\bar{B}^0_s: & M_{Z_2}>10.75\; \mbox{TeV}, \\ \label{lowerb}
D^0-\bar{D}^0: & M_{Z_2}>0.18\; \mbox{TeV}.  \end{eqnarray}

This shows that the strongest constraint comes from the $B^0_d-\bar{B}^0_d$ system, which raises the lower bound on $M_{Z_2}$ to a value larger than $11.54$ TeV, as compared with the bound in Eq.~(\ref{bounds}). This in turn reduces the allowed range of values for $\theta$ to: $-0.00006 \leq \theta \leq 0.00016$.

It can be easily checked that in order to keep low bounds on $M_{Z_2}$ in Eq.~(\ref{lowerb}), the third family of quarks must transform differently, that is, must be assigned to the 4-plet. It must also be stressed that these new bounds depend both on the assumed contribution of the $Z_2$ exchange to the mass differences and on the particular parametrization of the CKM matrix adopted. Thus, they can only be considered as a roughly estimate of the size of the constraints imposed by neutral meson mixing.

\subsection{\label{sec:sec5b}Oblique corrections}
The 3-4-1 extension of the SM presented here predicts new heavy particles: new exotic up-type quarks $U_i$, $U^\prime_i$ ($i=1,2$) and down type quarks $D_3$, $D^\prime_3$, new exotic leptons $E_\alpha$, $E^\prime_\alpha$ ($\alpha=1,2,3$), new gauge bosons $K^{\pm}$, $K^0\;(K^{\prime 0})$, $X^{\pm}$, $X^0\;(X^{\prime 0})$, $Y^0\;(Y^{\prime 0})$, $Z^{\prime\prime}$, and $Z^\prime$, and extra charged and neutral scalars. Provided these new particles feel the electroweak interaction, they should give corrections to the electroweak precision measurements through their effects on the $W$ and $Z$ vacuum polarization amplitudes. Such ``oblique" corrections are usually expressed in terms of the $S$, $T$ and $U$ parameters at the one loop level \cite{peskin}. Since current fits to electroweak precision observables show that $S$ and $T$ are small negative numbers and $U$ is close to zero \cite{pdg}, these parameters can be optimally used to constraint the new physics model building. So, we should determine whether additional constraints on the parameter space of the 3-4-1 model arise from examination of the oblique corrections or not. Eventhough a detailed calculation of $S$, $T$ and $U$ in the 3-4-1 model is out of the scope of the present work, several comments are in order.

First, note that in the limit $v = v'$ and $V = V'$, which we have assumed through this work, the new neutral gauge boson $Z^{\prime\prime}$ decouples from the low energy physics and, consequently, does not contribute to the oblique corrections which are only sensitive to $SU(2)$ breaking. $Z^\prime$ does not contribute because enters only at tree level through $Z-Z^\prime$ mixing whose effect on the $\rho$ parameter (which is equivalent to $T$) was included in the fit to $Z$-pole observables in the previuos section.

Since all the new quarks and leptons are $SU(2)$ singlets they do not contribute to $S$, $T$ and $U$. With regard the remaining new gauge bosons in Eq.~(\ref{chmass}), we can see that $Y^0\;(Y^{\prime 0})$ does not contribute because is also a $SU(2)$ singlet. The $SU(2)$ doublets $(K^+,K^0)$ and $(X^+,X^0)$ (and their conjugates) will contribute as long as there is mass splitting between the members of each doublet. As it is clear from Eq.~(\ref{chmass}), in the general case $v \ne v'$ and $V \ne V'$ the symmetry breaking pattern in Eq.~(\ref{break}) induces this mass splitting, however, in the particular case $v = v',\,V = V'$ we are considering, the members of each $SU(2)$ doublet are degenarate in mass and their contribution to the oblique corrections vanish. Thus, in the limit $v = v',\,V = V'$ only new physical Higgs fields will contribute radiatively to $S$, $T$ and $U$ and will be important in order to further constraint the parameter space of the 3-4-1 model.

The complete calculation of the scalar contributions to the oblique parameters requires the diagonalization of the full scalar sector of the theory in order to identify both the Goldstone bosons and the physical scalar fields. Since the four Higgs 4-plets in Eq.~(\ref{scalars}) contain a total of 32 degrees of freedom and in spite of the simplifications introduced by the $Z_2$ discrete symmetry, the diagonalization of the scalar sector is not a simple task in the context of the 3-4-1 model. Notwithstanding, the following general considerations can be done. 

After the symmetry breaking, the original 32 degrees of freedom become 15 Goldstone bosons (6 electrically charged and 9 neutral) which are swallowed up by the 15 massive gauge bosons in the model. So, we are left with 17 physical Higgs particles. With the hierarchy $V \sim V^\prime >> v \sim v^\prime$ and ignoring $Z-Z^\prime$ mixing, the $SU(2)$ doublets $(\phi^0_1,\phi^+_1)$ (coming from $\phi_1$) and $(\phi^-_2,\phi^0_2)$ (coming from $\phi_2$) form a standard THDM with $\tan\beta = v/v^\prime$ for which the oblique corrections have been studied in detail \cite{thdm2}. The THDM contains five physical Higgs fields $H^\pm$, $H^0_{1,2}$ and $A^0$ \cite{thdm1}. The remaining 12 scalar particles (four charged and eight neutral) are, in general, mixtures of $SU(2)$ doublets and singlet scalars with a small mixing angle $\alpha$ which, in the limit $v = v',\,V = V'$, is given by $\tan \alpha = v/V$. Contributions of these heavy scalars to the oblique parameters depend on this mixing and, as a result, they are suppressed by the square $\sin^2 \alpha \approx \tan^2 \alpha \approx \alpha^2$ of the small mixing angle. However, they also depend on the mass splitting between extra Higgs fields and on the mass splitting between extra scalars and extra gauge bosons different from $Z^{\prime\prime}$ and $Z^\prime$ (see, for example, the analysis for the 3-3-1 model presented in Ref.~\cite{Liu-Ng}). So, we can expect that in the case $v = v',\,V = V'$ in which the extra gauge bosons are degenerate, and the limit of small Higgs mass splitting (which should be ensured by the symmetry breaking hierarchy $V \gg v$), the scalar contributions should be within the room allowed by the current small values of the $S$, $T$ and $U$ parameters. 

\section{\label{sec:sec6}Confronting 3-4-1 three-family models without exotic electric charges}
The review of 3-4-1 models presented in Sec.~\ref{sec:sec2} shows that the restriction to particles without exotic electric charges picks up two class of anomaly free models characterized, respectively, by the values $b=c=1$ and $b=1,\;c=-2$ for the parameters in the electric charge generator in Eq.~(\ref{ch}). Two important differences between the two class of models can be recognized
which will enable us to perform a comparison of their predictions. 

First, in models for which $b=c=1$ (as the model analyzed in this paper), anomaly cancellation between the families implies not only a family of quarks transforming different from the other two but also a non-universal hypercharge $X$ for the left-handed quark multiplets. As a consequence, the couplings $g_{iV}$ and $g_{iA}$ ($i=1,2$) of fermions to the neutral currents $Z_1$ and $Z_2$, listed in tables 
\ref{tab3} and \ref{tab4} for Model B, are family dependent. Models for which $b=1,\;c=-2$ instead, have the particular feature that, notwithstanding one family of quarks transforms differently under the $SU(4)_L$ subgroup, the three families have the same hypercharge $X$ with respect to the $U(1)_X$ subgroup. So, the couplings of all the fermion fields to $Z_1$ and $Z_2$ are family-universal (for details see Ref.~\cite{spp}).

Second, eventhough in both class of models and in the limit $v^\prime=v,\; V^\prime=V$ the mixing between neutral currents can be contrained to occur between $Z$ and $Z^\prime$ only, in $b=c=1$ models the $Z^{\prime\prime}$ current couples only to exotic fermions and thus decouples completely from the low energy physics, and the left-handed couplings of $Z^\prime$ to the SM quarks, being flavor nondiagonal, induce tree level FCNC. By contrast, for models in the $b=1,\;c=-2$ class it is the $Z^{\prime\prime}$ current which couples nondiagonally to ordinary quarks thus leading to FCNC, while, due to the fact that the couplings of the fermion fields to $Z_1$ and $Z_2$ are family-universal, there are not FCNC transmitted by the $Z^\prime$ boson.

These differences affect the predictions of both class of models in two ways. In $b=c=1$ models, in which the couplings of fermion fields to $Z_1$ and $Z_2$ are family dependent, the issue of quark family discrimination  must be considered because affects the constraints on the parameter space $\theta-M_{Z_2}$ coming both from the fit to Z-pole observables and from FCNC processes. The analysis leads to the conclusion stated in the previous section according to which the third family of quarks must transform differently. Clearly, in $b=1,\;c=-2$ models, where there is not family dependence in the fermion couplings to $Z_1$ and $Z_2$, the effects of quark family nonuniversality are absent. On the other hand, since in $b=c=1$ models there are tree level FCNC mediated by the $Z^\prime$ boson, the constraints coming from FCNC data raise the lower bound on $M_{Z_2}$ obtained from the fit to Z-pole observables. In $b=1,\;c=-2$ models instead, $Z^\prime$ does not transmit FCNC and the lower bound on $M_{Z_2}$ is not affected by the constraints coming from the experimental results on FCNC (which, in this case, generate a lower bound on the $Z^{\prime\prime}$ mass). In addition, although both class of models predict a small $Z^\prime-Z$ mixing angle of the order of $10^{-4}$, $b=1,\;c=-2$ models predict a lower bound on the mass of the new neutral gauge boson $Z_2$ smaller than the one predicted by $b=c=1$ models. In fact, in $b=c=1$ models this bound is of the order of 2 TeV (see Eq.~(\ref{bounds}). See also Ref.~\cite{sgp}), while in $b=1,\;c=-2$ models this lower bound is found to be $M_{Z_2}\geq 890$ GeV \cite{newbound}, which is just below the TeV scale.
 
These considerations allows us to conclude that $b=1,\;c=-2$ models are favoured in the sense that they have a better chance to be tested in the upcoming generation of accelerators.

\section{Summary and conclusions}
We have presented an anomaly-free extension of the standard model of the electroweak interaction based on the gauge group $SU(4)_L\otimes U(1)_X$, which does not contain exotic electric charges. This last condition fixes the values $b=c=1$ or $b=1,\;c=-2$ for the parameters in the electric charge generator in Eq.~(\ref{ch}). Model B in Ref.~\cite{sp}, for which $b=c=1$, has not been previously studied in the literature and it has been considered here. A similar analysis for a model with $b=1,\;c=-2$ was done in Ref.~\cite{spp}, but the issue of quark family discrimination and the constraints imposed by FCNC effects were not discussed there.

In the model, anomaly cancellation preserves universality in the lepton sector, but forces one family of quarks to transform differently from the other two. As a consequence, FCNC arise.

The breaking of the 3-4-1 symmetry to the SM is achieved with a set of four $4^*$-plet of Higgs scalars, which set two different mass scales: $V\sim V^\prime >> \sqrt{v^2+v^{\prime 2}} =174$ GeV, with $v\sim v^\prime$. The imposed alignment of the vacuum, combined with an anomaly-free discrete $Z_2$ symmetry, produces a simple mass splitting between ordinary and exotic charged fermions, in such a way that all the ordinary quarks and charged leptons acquire masses at the energy scale $v$ and, separately, the exotic fermions acquire masses at the high energy scale $V$. Since the $Z_2$ symmetry forbids mixing between ordinary and exotic quarks, unitarity violation of the CKM matrix is absent in the model. 
 
The neutral leptons remain massless after the symmetry breaking. Notwithstanding, this sector can be easily extended by adding the neutral singlets required to accomodate neutrino phenomenology. 

In the charged gauge boson sector, the alignment of the vacuum also avoids mixing of the lightest mass eigenstate $W^{\pm}_\mu$ in Eq.~(\ref{chmass}), which we identify with the charged gauge boson of the SM. 

In the neutral gauge boson sector, instead, the three neutral currents $Z_\mu$, $Z'_\mu$ and $Z''_\mu$ predicted by the model, mix up. The current $Z''_\mu$, however, decouples from the low energy physics if we assume $v' = v$ and $V' = V$. It remains a mixing between $Z_\mu$ and $Z'_\mu$ which, by using LEP and SLC data at Z-pole and APV data, is constrained to obey the bounds: $M_{Z_2} \geq 2.02$ TeV and $-0.00017 \leq \theta \leq 0.0007$, for the mass of the new neutral gauge boson and the $Z-Z'$ mixing angle, respectively. These bounds have been further constrained by using experimental data from neutral meson mixing in the analysis of FCNC effects associated to quark nonuniversality. Assuming  the Fritzsch ansatz for the quark mixing matrix and taking complex phases equal to zero, we have found that the strongest constraint comes from the $B^0_d-\bar{B}^0_d$ system, which reduces the parameter space to: $M_{Z_2}> 11.5$ TeV and $-0.00006 \leq \theta \leq 0.00016$. The inclusion of CP-violating phases must not substantially weakening these bounds. It must be however recognized that the bounds from neutral meson  mixing are always obscured by the lack of knowledge of the mixing matrix entries.

Eventhough we have adopted the Fritzsch ansatz, another textures for the quark mixing matrix can also be used. Although the numerical values in Eq.~(\ref{lowerb}) might be modified, it is to be expected that the general outcome will be the same: the reduction of the region in the parameter space allowed by the fit to Z-pole observables. This is not surprising since FCNC through $Z_2$ exchange occur already at tree level in the model, while electroweak precision observables are associated to loop level processes. 

Another possible analysis, which does not require any specific parametrization of the CKM matrix, would be to use the bounds in Eq.~(\ref{bounds}) in order to constraint the size of the mixing matrix elements \cite{jars}. This type of approach requires a more careful treatment and it is outside the scope of the present work.

For the study of the model we have picked out the heaviest family of quarks to be the one transforming differently under the gauge group. In the main text of the paper we have shown that this must be the case if we want to end with a 3-4-1 scale of the order of a few TeV and, consequently, with a model able to be tested in the upcoming generation of accelerators. 

We have also discussed the constraints coming from the contribution of the exotic particles in the model to the one-loop oblique parameters $S$, $T$ and $U$. Extra heavy fermions do not contribute because all of them are $SU(2)$ singlets. In the particular case $v' = v,\;V' = V$ (which we have assumed through the paper), the extra gauge bosons masses are equal and their contribution vanish. So, in this case, the only contribution comes from the extra heavy scalars. On account of the complexity of the scalar sector in our model (four Higgs 4-plets with 32 degrees of freedom), a complete calculation of these contributions will be postponed to a future work.

Finally, a comparison of the predictions from the two classes of 3-4-1 models without exotic electric charges shows that models for which $b=1,\; c=-2$ are preferred in the sense that they give a lower bound on the mass of the new neutral gauge boson $Z_2$ smaller than the one predicted by models with $b=c=1$. Moreover, different from $b=c=1$ models, the lower bound $M_2\geq 0.89$ TeV obtained in $b=1,\; c=-2$ models is not affected by the constraints coming from FCNC data and is just below the TeV scale, which means that this class of models have a better chance to be tested at the forthcoming CERN LHC facility.  

\section*{ACKNOWLEDGMENTS}
L.A.S. acknowledge financial support from DIME at Universidad Nacional de Colombia-Sede Medell\'\i n. L.A.W-T acknowledge finacial support from American University of Sharjah. J.I.Z. acknowledge partial financial support from CODI at Universidad de Antioquia.

\begin{thebibliography}{}
\bibitem[34]{newbound} 
Here we must point out that on account of an erroneous calculation of the term $\Delta Q^\prime_W$ in the fit to Z-pole observables, the bound $0.67\;\mbox{TeV}\leq M_{Z_2} \leq 6.1\;\mbox{TeV}$ obtained in Ref.~\cite{spp} is incorrect. The correct calculation gives a fit which produces only the lower bound $M_{Z_2}\geq 0.89$ TeV, with no upper bound present.
\end{thebibliography}
\end{document}